\documentclass[aps,prb,twocolumn]{revtex4}
\usepackage{graphicx}% Include figure files
\usepackage{dcolumn}% Align table columns on decimal point
\usepackage{bm}% bold math
\usepackage{multirow}
\usepackage{hyperref}
\usepackage{mathtools}
\usepackage{braket}
\usepackage{color}
\usepackage{url}
\usepackage{epstopdf}
\usepackage{amssymb}
\usepackage{amsmath}
\begin{document}
\preprint{APS/123-QED}
%\title{CeAgAs$_2$ :  A Kondo lattice system exhibiting mixed valence and magnetic ordering}
\title{Magnetocrystalline anisotropy in  the Kondo lattice  compound CeAgAs$_2$ }%\title{Magnetocrystalline anisotropy in CeAgAs$_2$ single crystal}
%\author{Pranab Kumar Das, Amitava Bhattacharyya, S. K. Dhar, Ruta Kulkarni and A. Thamizhavel}
\author{Rajib Mondal}
\affiliation{Department of Condensed Matter Physics and Materials Science, Tata Institute of Fundamental Research, Homi Bhabha Road, Colaba, Mumbai 400 005, India.}

\author{Rudheer Bapat}
\affiliation{Department of Condensed Matter Physics and Materials Science, Tata Institute of Fundamental Research, Homi Bhabha Road, Colaba, Mumbai 400 005, India.}

%\author{Amitava Bhattacharyya}
%\affiliation{Rutherford Appleton Laboratory, Harwell Oxford, Didcot, OX11 0QX}

%\author{Ruta Kulkarni}
%\affiliation{Department of Condensed Matter Physics and Materials Science, Tata Institute of Fundamental Research, Homi Bhabha Road, Colaba, Mumbai 400 005, India.}

\author{S. K. Dhar}
\affiliation{Department of Condensed Matter Physics and Materials Science, Tata Institute of Fundamental Research, Homi Bhabha Road, Colaba, Mumbai 400 005, India.}

\author{A. Thamizhavel}
\affiliation{Department of Condensed Matter Physics and Materials Science, Tata Institute of Fundamental Research, Homi Bhabha Road, Colaba, Mumbai 400 005, India.}

\date{\today}

\begin{abstract}

We report on the single crystal growth and anisotropic physical properties of CeAgAs$_2$.  The compound crystallizes as on ordered variant of the HfCuSi$_2$-type crystal structure and adopts the orthorhombic space group $Pmca$~(\#57) with two symmetry inequivalent cerium atomic positions in the unit cell.  The orthorhombic crystal structure of our single crystal was confirmed from the powder x-ray diffraction and from electron diffraction patterns obtained from the transmission electron microscope.   The anisotropic physical properties have been investigated on a good quality single crystal by measuring the magnetic susceptibility, isothermal magnetization, electrical transport and heat capacity.  The magnetic susceptibility and magnetization measurements revealed that this compound orders antiferromagnetically with two closely spaced magnetic transitions  at $T_{\rm N1} = 6$~K and $T_{\rm N2} = 4.9$~K.  Magnetization studies have revealed a large  magnetocrystalline anisotropy  due to the crystalline electric field (CEF) with an easy axis of magnetization along the [010] direction.  The magnetic susceptibility measured along the [001] direction exhibited a broad hump in the temperature range 50 to 250~K, while typical Curie-Weiss behaviour was observed along the other two orthogonal directions. The electrical resistivity and the heat capacity measurements revealed that CeAgAs$_2$ is a Kondo lattice system with a magnetic ground state.  

\end{abstract}

%\pacs{81.10.-h, 71.70.Ch, 75.10.Dg, 75.50.Gg, 71.70.Gm, 75.30.Sg} 
\maketitle

\section{Introduction}
One of the most widely investigated topics of research in  condensed matter physics during the last few decades is the competition between the Ruderman, Kittel, Kasuya and Yosida (RKKY) interaction and the Kondo effect in Ce and Yb based intermetallic compounds.  The investigations made on these compounds have discovered a plethora of new phenomena like heavy fermion, valence fluctuation, superconductivity, quantum phase transition etc~\cite{Fisk, Lawrence, Onuki, Settai, Gegenwart, Flouquet}.  In recent years there are many reports on Ce-based compounds which can be tuned to quantum critical point (QCP) with pressure as the tuning parameter~\cite{Weng, Sugitani, Onuki2}.   All these interesting physical properties are due to the result of the  hybridization of the localized $4f$ electron  with the conduction electrons.  A weak hybridization usually results in a magnetic ground state while a stronger hybridization results in a  weaker localization of the $4f$ electron pushing the system eventually to a non-magnetic intermediate valence state.  The hybridization strength and the effects due to the crystal electric field depend on the atomic arrangement of the environment around the Ce atom in the unit cell, which includes the inter-atomic distances between the Ce atom and other atoms and ligands on different crystallographic sites. Most of the cerium based compounds possess a single unique atomic position for the Ce-atom in the unit cell. However, there are a number of cerium compounds which have multiple Ce atomic sites in the unit cell~\cite{Gschneidner}, like for example, Ce$_7$Ni$_3$~\cite{Umeo}, Ce$_2$PdGe$_3$~\cite{Baumbach}, Ce$_4$Co$_2$Sn$_5$~\cite{Pani}, CeRuSn$_3$~\cite{Anand}, CeRhSn$_3$~\cite{Anand2}, Ce$_2$Rh$_3$Ge~\cite{Falkowski}, Ce$_2$RuZn$_4$~\cite{Eyert} and CeRuSn~\cite{Fik} etc.  The compounds with multiple Ce-sites have been mostly studied in the polycrystalline form.  Here we report on the  magnetic properties of a single crystal sample of CeAgAs$_2$, which allows us to study the anisotropic behavior in detail.  

The compounds of the form Ce$T$X$_2$, where $T$ is a transition metal and $X$ is either Sb or As, crystallize mostly in the HfCuSi$_2$-type crystal structure with the space group $P4/nmm$ (\#129) that hosts a unique atomic position for Ce atom at the $2c$ site.  The anisotorpic physical properties of Ce$T$Sb$_2$ have been studied quite extensively~\cite{Myers, Takeuchi, Thamizh}.  For example CeAgSb$_2$ is a ferromagnet with the easy axis of magnetization along the [001]-direction and a saturation moment of only 0.45~$\mu_{\rm B}$/Ce, which remains constant upto fields as high as 45~T.  On the other hand, the isostructural CeAuSb$_2$  becomes an antiferromagnet at 6~K and exhibits field induced quantum criticality~\cite{Balicas}. In view of the wide variety of magnetic properties exhibited by the antimonides, the arsenides of this series of compounds have also been synthesized and studied for their crystal structure and magnetism. Owing to the huge vapor pressure of arsenic,  the compounds of Ce$T$As$_2$ have been synthesized in polycrystalline form~\cite{Demchyna, Maria, Eschen, Doert}.  In the present work, we were successful in growing the single crystals of CeAgAs$_2$ using the flux method.  Here we report the interesting anisotropic magnetic properties exhibited by this compound.

\section{Experiment}

We have employed the high temperature solution growth or the so called flux growth method to grow the single crystals of CeAgAs$_2$.  Typically, for flux growth one of the constituents that has the lowest melting point is used as a flux (solvent) to grow the single crystal when supersaturation is achieved on cooling.  But in CeAgAs$_2$ none of the elements possess a low melting point and hence we have used the eutectic composition of Ag-As, as a flux to grow the single crystals~\cite{Baren}.  The binary phase diagram of Ag-As reveals a eutectic composition at 75:25 atomic percent of Ag:As, with a low melting point of 540~$^{\circ}$C.  The starting materials of high purity Ce (99.9\% Ames Lab), Ag (99.99\%) and As(99.999\%) were taken in the ratio 1 : 37 : 14, which took into account the eutectic composition of the excess Ag:As flux.  The elements were placed in a high quality recrystallized alumina crucible and subsequently sealed in a quartz ampoule under partial pressure of argon gas.  The ampoule was loaded in a resistive heating box type furnace and the temperature of the furnace was raised at the rate of 15~$^{\circ}$C/h to 1050~$^{\circ}$C and held there for about 24 h for homogenization of solution.  Then the  the furnace was cooled to 700~$^{\circ}$C at the rate of 2~$^{\circ}$C/h at which point the ampoule was centrifuged to remove the flux.  Large, shiny, flat platelet like single crystals with typical size of 10 x 6 x 0.3 mm$^3$ were obtained.  The flat plane of the crystal corresponds to the (001)-plane.  Powder x-ray, Laue diffraction and transmission electron microscopy (TEM) was performed on a few grown crystals to know the phase purity and the crystallographic orientation and the superstructure nature of the crystal.  The crystal was cut into a bar shaped sample using the spark erosion cutting machine for the anisotropic physical property measurements.  The electrical resistivity, heat capacity and  magnetization  measurements were performed in a physical property measurement system (PPMS) and magnetic property measurement system (MPMS) from Quantum Design, USA, respectively.

\section{Experimental Results}
\subsection{X-ray diffraction}

\begin{figure}[!]
\includegraphics[width=0.45\textwidth]{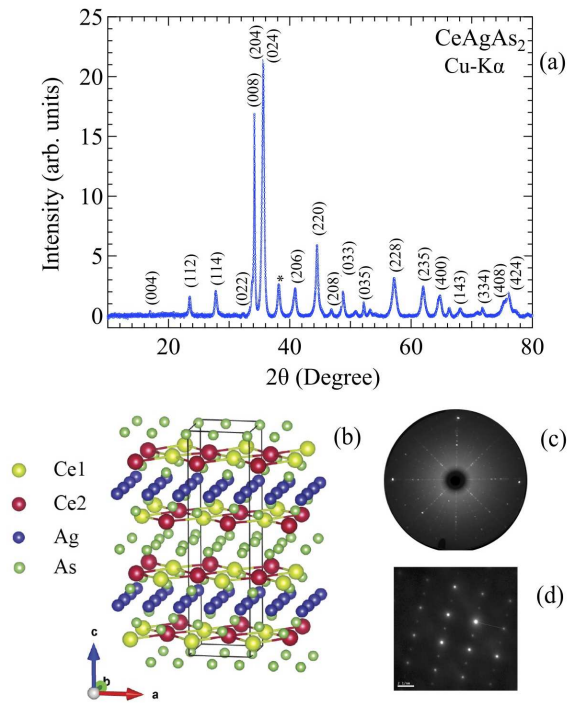}
\caption{\label{Fig1}(Color online) Powder x-ray diffraction pattern of CeAgAs$_2$    (b) the crystal structure of CeAgAs$_2$, (c)  The Laue diffraction pattern corresponding to (001) plane of the crystal and (d) TEM diffraction pattern depicting the superstructure spots.}
\end{figure}

Since the crystals were grown completely from an off-stoichiometric melt, at first we performed a powder x-ray diffraction (XRD) to confirm the phase purity.  Figure~\ref{Fig1} shows the powder x-ray diffraction pattern of our sample. Here it may be pertinent to first recall that a  detailed crystal structure analysis of CeAgAs$_2$ has been previously performed by Demchyana \textit{et al}~\cite{Demchyna} which showed  that unlike the other members of the Ce$T$X$_2$ ($T$ = Transition metal; $X$ = Sb or As) CeAgAs$_2$ does not adopt the tetragonal HfCuSi$_2$ type crystal structure.   It crystallizes in the orthorhombic crystal structure with the space group $Pmca$ ($\#57$) with lattice parameters $a = 5.7441$~\AA, $b = 5.7696$~\AA~ and $c = 21.0074$~\AA, which are related to the tetragonal HfCuSi$_2$ structure by $\sqrt{2} \times a_{\rm T}$ and $2 \times c_{\rm T}$.  
%*********************FIGURE 1********************************
%****************************************************************
The \textit{Pmca} space group has two atomic positions for Ce atoms, Ce1 and Ce2 at $4d$ positions.  The crystal structure of CeAgAs$_2$ is shown in Fig.~\ref{Fig1}(b).  The nearest distance between the Ce1-Ce1 and Ce2-Ce2 atoms is $5.758$~\AA~ while that of Ce1-Ce2 is $4.078$~\AA.   The Laue diffraction pattern of the flat plane of the as grown crystal is shown in Fig.~\ref{Fig1}(c), which corresponds to the (001)-plane of the crystal.  Well defined spots with clear symmetry pattern confirms the good quality of the grown single crystal.  Furthermore, the superstructure nature of the unit cell along the $c$-axis is observed using the transmission electron microscope (TEM) and the observed diffraction pattern is shown in Fig.~\ref{Fig1}(d).  The diffused spots confirms the superstructure nature of the unit cell while this is not observed in the TEM diffraction pattern of CeCuAs$_2$, not shown here for brevity.

\subsection{Magnetic susceptibility and magnetization}

%*********************FIGURE 1********************************
\begin{figure}[!]
\includegraphics[width=0.45\textwidth]{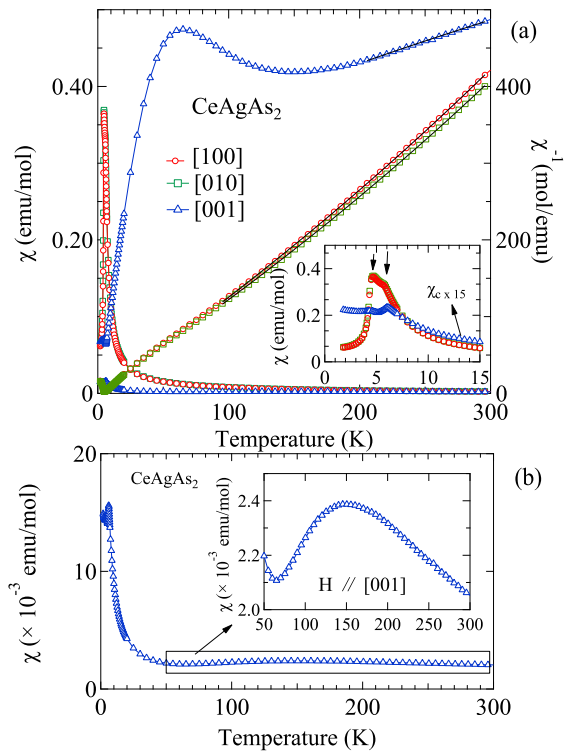}
\caption{\label{Fig2}(Color online) (a)  Temperature dependence of anisotropic magnetic susceptibility along the three principal crystallographic directions.  A huge drop in the magnetic susceptibility along [100] and [010] directions is observed.  Also shown is the inverse magnetic susceptibility and the modified Curie-Weiss fit.  Inset shows two clear magnetic transitions as indicated by the arrow.  (b)  Zoomed in view of the magnetic susceptibility along [001] direction in the temperature range 50 - 300~K. }
\end{figure}
%****************************************************************

The temperature dependence of the \textit{dc} magnetic susceptibility measured along the three principal crystallographic directions from 1.8 to 300~K,  is shown in the main panel of Fig.~\ref{Fig2}(a).  The magnetic susceptibility along the [100] and [010] directions increases more rapidly below about 50~K and then exhibits a small cusp at 6~K followed by a sharp drop at 4.9~K indicating two magnetic transitions (see the inset of Fig.~\ref{Fig2}(a)). From the magnetization measurements \textit{vide infra}, and from the previous neutron diffraction studies~\cite{Doert} on a polycrystalline sample, the two magnetic transitions are confirmed to be antiferromagnetic in nature.   Just below the first antiferromagnetic transition at $T_{\rm N1} = 6$~K, the magnetic susceptibility increases and then it drops down more rapidly below $T_{\rm N2} = 4.9$~K.  The  drop in the magnetic susceptibility  ascertains the antiferromagnetic nature of the transition at 4.9~K and confirms the easy axis of magnetization as [100] or [010].  For $H~\parallel~[001]$, at $T_{\rm N1}$ the susceptibility shows a drop followed by a small rise at the second transition at $T_{\rm N2}$, being virtually temperature independent at lower temperatures suggesting [001] as the hard axis of magnetization.  The anisotropy within the $ab$ plane is weak while the anisotropy between the $ab$ plane and the $c$-axis is quite large.   The curvilinear behaviour of the inverse magnetic susceptibility along the [100] and [010] directions indicates the simple Curie-Weiss law cannot be used to analyse the magnetic susceptibility data.  Hence, we have employed the modified Curie-Weiss law: $\chi^{-1} = (\chi_0 + \frac{C}{(T-\theta_{\rm p})})^{-1}$ to fit the inverse magnetic susceptibility above 100~K for [100] and [010] directions.  Here $\chi_0$ is the temperature independent part of the magnetic susceptibility and C is the Curie constant.   We have obtained $\chi_{0} = -4.434 \times 10^{-4}$~emu/mol, an effective magnetic moment of 2.59~$\mu_{\rm B}/$Ce and the paramagnetic Weiss temperature  $\theta_{\rm p} = -0.20$~K for $H~\parallel$~[100] and $\chi_{0} = -4.797~\times 10^{-4}$~emu/mol, $\mu_{\rm eff} = 2.66~\mu_{\rm B}/$Ce and  $\theta_{\rm p} = -2.32$~K for $H~\parallel$~[010] direction. The obtained effective magnetic moment values are very close to the theoretical value of trivalent Ce, suggesting a local moment behaviour. The modified Curie-Weiss law  could not result in a good fitting for $H~\parallel$~[001] direction.  However, if we fix the effective magnetic moment to $2.54~\mu_{\rm B}$/Ce a reasonably good fitting is obtained in the temperature range 250 - 300~K with $\chi_{0} = 7.079~\times 10^{-4}$~emu/mol and $\theta_{\rm p} = -299$~K as shown in the main panel of Fig~\ref{Fig2}(a).  Figure~\ref{Fig2}(b) shows the magnetic susceptibility for $H~\parallel$~[001] direction, which is very weak (of the order of 10$^{-3}$emu/mol).  It is obvious from the susceptibility data that the Curie-Weiss behaviour is absent and a broad hump is observed in the temperature range 50 to 300~K.  This type of behaviour is usually attributed to mixed valent or intermediate valent character~\cite{Falkowski, Adroja}. However, in the present case, we shall see that it arises due to a strong crystal field effect.

%*********************FIGURE 3********************************
\begin{figure}[!]
\includegraphics[width=0.45\textwidth]{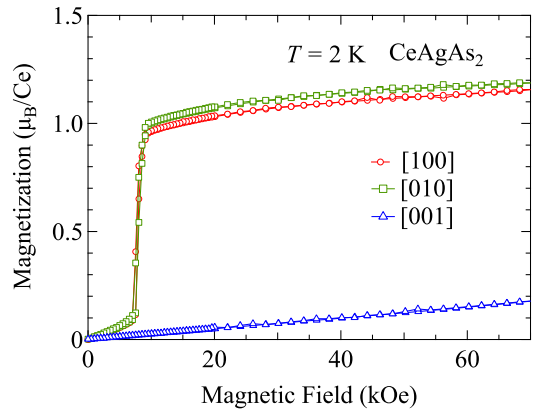}
\caption{\label{Fig3}(Color online) Isothermal magnetization measured at $T=2$~K along the three principal crystallographic directions.  A sharp first order like metamagnetic transition is observed at 7.5~kOe for $H~\parallel$~[100] and [010] directions. }
\end{figure}
%****************************************************************

The isothermal magnetization measured at $T=2$~K is shown in Fig.~\ref{Fig3}. In conformity with the magnetic susceptibility data presented above, the isothermal magnetization exhibits an easy plane magnetocrystalline anisotropy.  A very sharp spin flip like metamagnetic transition is observed at field of 7.5~kOe for $H~\parallel$~[100] and [010] directions, and for fields greater than 10~kOe, the magnetization almost saturates.  The saturation value of the magnetization is about 1.2~$\mu_{\rm B}$/Ce, which is much less than that expected for the free-ion value of Ce in its trivalent state ($g_{\rm J} J = 2.14~\mu_{\rm B}$/Ce).  On the  other hand the magnetization along  [001] increases very gradually and attains a value of only 0.18~$\mu_{\rm B}$ at 70~kOe,  thus confirming it as the hard axis of magnetization.  The low saturation value of the magnetic moment for fields as high as 70~kOe suggests a strong evidence for crystal electric field (CEF) and Kondo effect.  We have also performed the temperature dependence of isothermal magnetization and found that the critical field at which the metamagnetic transition appears shifts to lower fields and broadens as the temperature is increased and finally for temperature larger than $T_{\rm  N1}$ the magnetization shows an almost linear dependence on field in  the paramagnetic state, as expected in antiferromagnetic compounds.

\subsection{Electrical Resistivity}

\begin{figure}[b]
\includegraphics[width=0.5\textwidth]{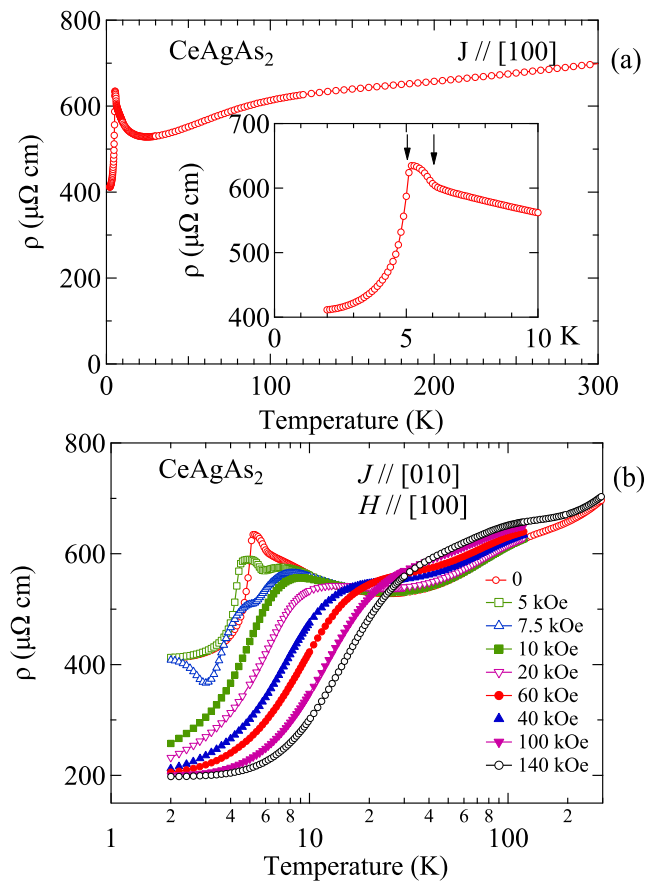}
\caption{\label{Fig4}(Color online) (a) Temperature dependence of electrical resistivity in the temperature range 1.8 - 300~K.  Inset shows the low temperature part, where the two magnetic transitions are clearly seen.  (b)  Magentic field dependence of electrical resistivity for $J~\parallel$~[100] and $H~\parallel$~[100] direction.   }
\end{figure}

The temperature dependence of electrical resistivity of CeAgAs$_2$ for current parallel to [100]-direction or $(ab)$-plane is shown in the main panel of Fig.~\ref{Fig4}.  We could not measure the resistivity for $J~\parallel$~[001] because of the small thickness of the sample.  At room temperature the electrical resistivity reads a value of about $700~\mu \Omega \cdot$cm which is quite large.  As the temperature decreases the resistivity decreases gradually and exhibits a very broad maximum centered around 100~K, which may be attributed to the thermal depopulation of the CEF levels.   Below 100~K, the resistivity falls off more rapidly showing a broad minimum around $25$~K and then increases  gradually. The resistivity data in zero field show a linear logarithmic temperature dependence in the range 20 to 6~K suggesting that CeAgAs$_2$ is a Kondo lattice system [Fig.~\ref{Fig4}(b)]. At 6~K a sudden change of slope is observed (see the inset of Fig.~\ref{Fig4}(a)) and the electrical resistivity increases more rapidly, which marks the $T_{\rm N1}$.  At $T_{\rm N2} = 4.9$~K, the electrical resistivity drops due to the reduction in the spin disorder scattering, thus confirming the two magnetic transitions observed from the magnetic susceptibility data.  The increase in the electrical resistivity just below $T_{\rm N1}$ is typical of compounds that exhibit superzone gap~\cite{Pranab_CeGe}, which occurs due to the  difference in the lattice periodicity and
%*********************FIGURE 4********************************
%**************************************************************** 
magnetic periodicity.  But here in the present case, the presence of superzone gap is ruled out because the previous neutron diffraction studies have revealed the propagation vector $k = [0,0,0]$~\cite{Doert}.  Hence the increase in the electrical resistivity in the region $T_{\rm N2} < T < T_{\rm N1}$ is mainly attributed to the spin fluctuation effect presumably due to the isosceles triangular arrangements of the Ce1 and Ce2 atoms along the [100]-direction.  It should be noted that the electrical resistivity begins to drop at $T_{\rm N2}$ from a value of $634~\mu \Omega \cdot$cm to about $411~\mu \Omega \cdot$cm at 2~K the lowest temperature measured.  The huge value of the residual resistivity is attributed to the spin fluctuations as discussed below.    The upturn in the resistivity below 25~K is reminiscent of a Kondo lattice system.  
 
We have also measured the electrical resistivity under applied magnetic fields and the plots are shown in Fig.~\ref{Fig4}(b).  The magnetic field is applied along the easy axis of magnetization direction.  With the increase in the magnetic field, the two magnetic transition temperatures decrease  and are not discernible for fields greater than 10~kOe down to 2~K.  This type of behaviour is generally observed in antiferromagnetic systems.  In an applied field of 7.5~kOe, the electrical resistivity shows an upturn at 3~K and the increasing trend is seen down to 2~K.  This anomalous behaviour of the electrical resistivity may be attributed to the spin flip like metamagnetic transition observed in the isothermal magnetization at 7.5~kOe, it is observed only at this field and  not  at higher or lower magnetic fields.   It is interesting to note that the residual resistivity decreases substantially with increasing magnetic fields suggesting a significant magnetic contribution due to spin fluctuations which are suppressed by applied magnetic fields.

\begin{figure}[!]
\includegraphics[width=0.5\textwidth]{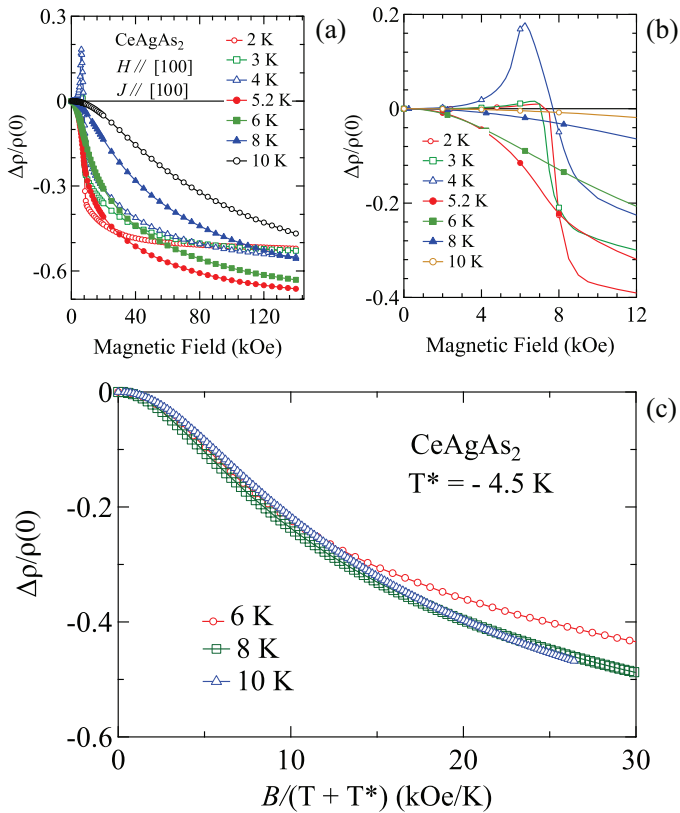}
\caption{\label{Fig5}(Color online) (a) Isothermal magnetoresistance $MR = \frac{\Delta\rho}{\rho(0)} = \frac{\rho(B) - \rho(0)}{\rho(0)}$ for CeAgAs$_2$ as a function of applied magnetic field for $J~\parallel$~[100] and $H~\parallel$~[100].  (b) The low field part of the isothermal magnetoresistance with sparse markers for clarity and (c) normalized magnetoresistance in the paramagnetic region plotted as a function $B/(T +T^*)$ }
\end{figure}

The isothermal field dependent magnetoresistance is shown in Fig.~\ref{Fig5}(a) and the low field part as a zoomed in view, with sparse markers,  is shown in Fig.~\ref{Fig5}(b). The variation of magnetoresistance with field is in good correspondence with the magnetization data presented earlier. At the lowest temperature of 2~K, the magnetoresistance is slightly positive and a sudden change in magnitude is observed in the negative direction at 7.5~kOe, suggesting the field induced ferromagnetic state.  A similar behaviour is observed for 3~K data as well.  At $T=4$~K, the magnetoresistance exhibits a sharp peak at 6.3~kOe and becomes negative at 7.5~kOe.  This may be attributed to the effect of field on spin fluctuations at this temperature.  In the paramagnetic region, the normalized magnetoresistance can be mapped on to a single curve if plotted as a function of $B/(T+T^*)$, where $T^*$ is the characteristic temperature, often identified as the Kondo temperature~\cite{Schlottmann}.  Our experimental data taken at selected temperatures can be mapped on to a single curve with a characteristic temperature $T^*$ of about -4.5~K. Negative $T^*$ value has been obtained for several antiferromagnetic Kondo lattice compounds~\cite{Pikul1, Pikul2} and it was attributed to the ferromagnetic correlations.  From the magnetic measurements it is clear that CeAgAs$_2$ orders antiferromagnetically.  The ferromagnetic correlations in this compound arises due to the parallel  alignment of magnetic moments in the layers of the $ab$ plane which are antiparallel along the $c$-axis as inferred from the neutron diffraction studies by  Doert \textit{et al}~\cite{Doert}.

%*********************FIGURE 5********************************
%**************************************************************** 

\subsection{Specific heat }

%*********************FIGURE 6********************************
\begin{figure}[!]
\includegraphics[width=0.5\textwidth]{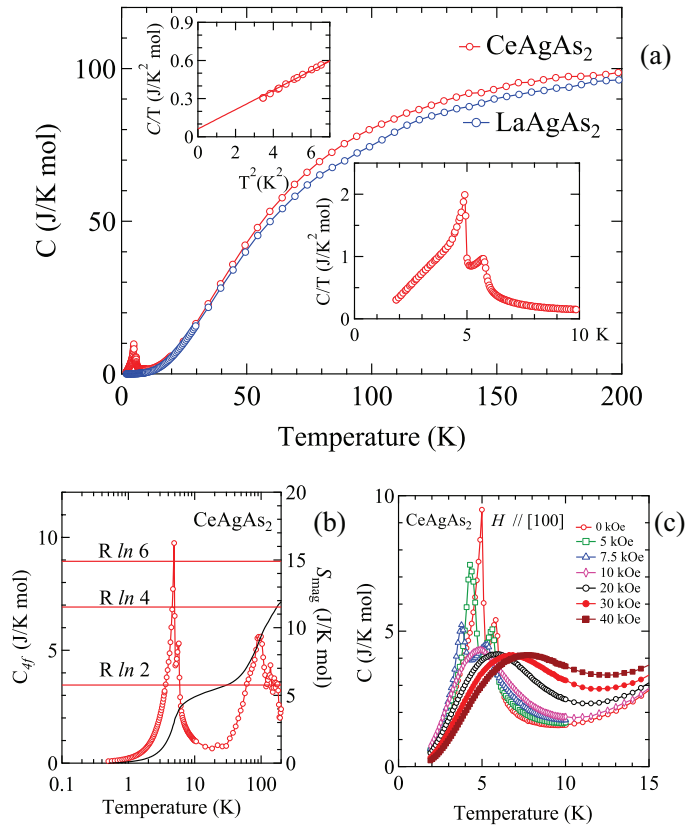}
\caption{\label{Fig6}(Color online) (a)Temperature dependence of heat capacity of CeAgAs$_2$ and the non-magnetic LaAgAs$_2$.  The upper inset shows the low temperature part of $C/T$ versus $T^2$.  The lower inset shows the low temperature part of specific heat in the form of $C/T$ versus $T$.  (b)  The magnetic part of heat capacity along with the entropy.  (c)  Field dependence of the heat capacity.}
\end{figure}
%**************************************************************** 

The temperature dependence of heat capacity of CeAgAs$_2$ and the non-magnetic reference LaAgAs$_2$ in the temperature range $2-200$~K is shown in the main panel of  Fig.~\ref{Fig6}(a). The heat capacity of CeAgAs$_2$ is larger than that of LaAgAs$_2$. The bulk nature of the magnetic transitions observed in magnetization and resistivity is confirmed by the two sharp peaks at $T_{\rm N1} = 6$~K and $T_{\rm N2}= 4.9$~K with appreciable magnitude (lower inset of Fig.~\ref{Fig6}(a)).  The low temperature data of LaAgAs$_2$ in the temperature range (2~ to $\sim 6$~K) was fitted to the expression $C = \gamma T + \beta T^3$, where the Sommerfeld coefficient $\gamma$ is the electronic contribution and $\beta$ is phonon contribution to the heat capacity.  The $\gamma$ and $\beta$ values thus obtained are  1.20~mJ/K$^2 \cdot$mol and $0.413$~mJ/K$^4 \cdot$mol, respectively.  The upper inset of Fig~\ref{Fig6}(a) shows the low temperature part of CeAgAs$_2$  in the form of $C/T$ versus $T^2$.  Similarly, we have obtained the $\gamma$ and $\beta$ values for CeAgAs$_2$ as 61.3~mJ/K$^2 \cdot$mol and 76.12~mJ/K$^4 \cdot$mol respectively. There is a substantial increase in $\gamma$ due to the Kondo effect. The enhanced value of $\beta$ is attributed to spin waves in CeAgAs$_2$, which follow a $T^3$ dependence in antiferromagnets. We have estimated the magnetic part of heat capacity by  the usual method of subtracting the heat capacity of LaAgAs$_2$ from that of CeAgAs$_2$.  The obtained $C_{\rm 4f}$ contribution to the heat capacity is shown in Fig.~\ref{Fig6}(b).  The magnetic entropy that is released at 6~K is only 70\% of $R$ln2 while the remaining 30~\% of entropy is released at around 30~K. This indicates that the ground state is a doublet ground state.   The  reduction in the magnetic entropy is mainly attributed to Kondo effect.  Similarly the jump in the magnetic part ($\Delta C_{\rm mag}$) of the heat capacity amounts to 9.02~J/K$\cdot$mol at $T_{\rm N1}$ and 4.53~J/K$\cdot$mol at  $T_{\rm N2}$.  The reduced jump in the heat capacity, compared to the value expected for $S = 1/2$ doublet state supports the existence of the Kondo effect in CeAgAs$_2$.

We have also performed the field dependence of  heat capacity of CeAgAs$_2$.  When the field was applied parallel to the flat plane of the crystal, there was no change in  heat capacity (not shown here for brevity).  Hence the crystals were aligned in such a way that the applied magnetic field is parallel to the easy axis of magnetization \textit{viz.}, [100].  Figure~\ref{Fig6}(c) shows magnetic field dependence of heat capacity at various applied magnetic fields.  It is evident from the figure that as the applied magnetic field is increased, the N\'{e}el temperatures shift towards lower temperature.  At a field of 7.5~kOe, where the metamagnetic jump is observed in the magnetization data, there is a sudden change in the magnitude of the heat capacity and for fields greater than 7.5~kOe, the heat capacity  shows only a broad hump which shifts towards the right side at higher fields.  At the lowest temperature measured, the in-field heat capacity is lower than its zero field value, which can be attributed to the breakdown of  Kondo coupling, between the localized $4f$ and conduction electrons with applied field~\cite{Pranab}. 

\section{Discussion}

The magnetic measurements along the three principal crystallographic directions revealed a strong anisotropy between the $ab$ plane and $c$-axis.  Furthermore, there is a strong suppression of the ordered moment and a reduced heat capacity jump together with a Schottky like peak in the $C_{\rm 4f}$ heat capacity caused by the CEF effect.  Hence,  we analysed the magnetocrystalline anisotropy based on the point charge model.  For the purpose of the CEF analysis, we have plotted the magnetic susceptibility in the form of $1/(\chi - \chi_0)$, where $\chi_0$ was determined from the modified Curie-Weiss fit as mentioned earlier. There are two Ce sites occupying the same $4d$ Wyckoff's position, with the same $x$ coordinate while $y$ and $z$ are different~\cite{Demchyna}. Assuming that the $4d$ site which hosts both the Ce-atoms has identical crystallographic environment, we have performed the CEF analysis.   The point symmetry of the $4d$ Wycoff's position is $m$ and hence possesses monoclinic site symmetry.  For monoclinic site symmetry, the $2J+1$ ground state of Ce-atom splits into three doublets.   In order to reduce the number of fitting parameters in the CEF analysis, we used the CEF Hamiltonian for the orthorhombic site symmetry which is given by,

%*********************FIGURE 8********************************
\begin{figure}[!]
\includegraphics[width=0.5\textwidth]{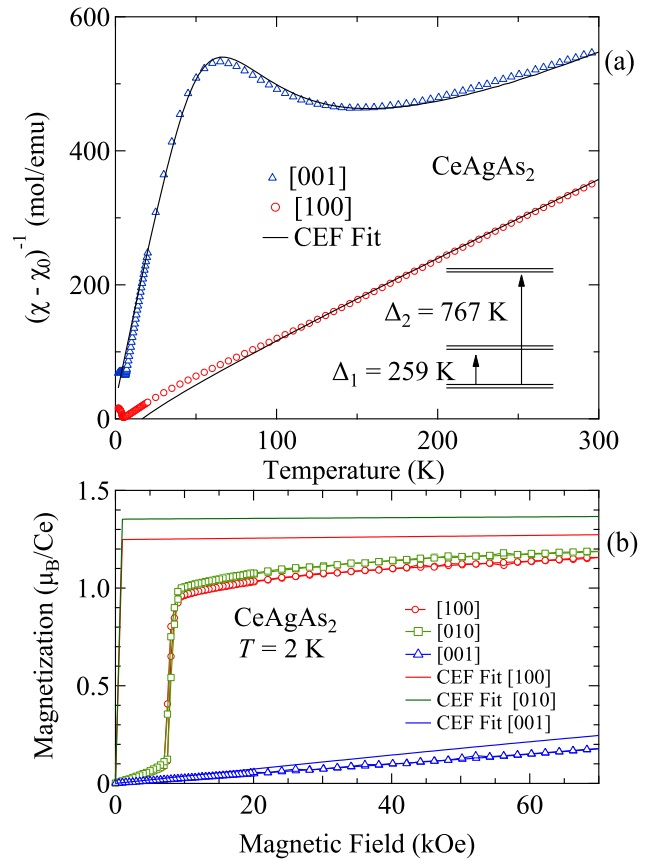}
\caption{\label{Fig8}(Color online) Temperature dependence of the inverse magnetic susceptibility plotted as $1/(\chi - \chi_0)$.  The solid lines in (a) are fits to Eqn.~\ref{eqn6}.  (b) CEF fits to the isothermal magnetization curves of CeAgAs$_2$ at 2~K along the principal crystallographic directions}
\end{figure}
%**************************************************************** 

\begin{equation}
\label{eqn5}
\mathcal{H}_{\rm CEF} = B_2^0{\bf O}_2^0 + B_2^2{\bf O}_4^2 + B_4^0{\bf O}_4^0 + B_4^2{\bf O}_2^2 + B_4^4{\bf O}_4^4,
\end{equation}

where $B_{\rm n}^{\rm m}$ are the CEF parameters and ${\bf O}_{\rm n}^{\rm m}$ are the Stevens operators~\cite{Stevens, Hutchings}.  In the above CEF Hamiltonian we have ignored the sixth order terms as they are zero for Ce atom.  The magnetic susceptibility including the molecular field contribution $\lambda$ is given by

\begin{equation}
\label{eqn6}
\chi^{-1} = \chi_{\rm CEF}^{-1} - \lambda_i,
\end{equation}

where $\chi_{\rm CEF}$ is CEF susceptibility.  The expression for the CEF susceptibility is given in our previous report~\cite{Pranab}.  We have also analysed the isothermal magnetization by using the following expression:

\begin{equation}
\label{eqn7}
M_{\rm i} = g_j \mu_{\rm B} \sum_n \braket{n | J_{\rm i} | n}  \frac{{\rm e}^{ -E_{\rm n}/k_{\rm B}T}}{\sum_n {\rm e}^{-E_{\rm n}/k_{\rm B}T}},
\end{equation}

where the $E_{\rm n}$ and the eigenfunction $|n>$ are determined by diagonalizing the total Hamiltonian

\begin{equation}
\mathcal{H = H_{\rm CEF}} - g_{\rm J} \mu_{\rm B} J_{\rm i}(H + \lambda_{\rm i} M_{\rm i}),
\end{equation}
where $\mathcal{H}_{\rm CEF}$ is given by Eq.~\ref{eqn5}, the second term is the Zeeman term and the third is the molecular field term.

In Fig.~\ref{Fig8}(a) we show the calculated CEF curves as solid lines, which reproduce reasonably the observed experimental susceptibility along the two principal crystallographic directions \textit{viz.} [100] and [001] direction.  We did not show the inverse susceptibility plot of [010] direction as it overlaps with the [100] direction. CEF analysis provides a goodfit to the experimental data. It nicely reproduces the broad peak in the susceptibility along the [001] direction. The CEF parameters and the energy levels thus obtained from the diagonalization of the CEF Hamiltonian are given in Table~\ref{table2}.   The $2J+1$ degenerate $J=5/2$ level is split into three doublets at energies 0, 259 and 767~K.  The ground state is predominantly  $\ket{\pm \frac{1}{2}}$  with the mixing from the $\ket{\pm \frac{5}{2}}$  and $\ket{\pm \frac{3}{2}}$   states.  The molecular field constant is highly anisotropic with positive values in the $ab$-plane along the $a$ and $b$ axes, while it is negative along the $c$-axis.  This is consistent with the alignment of the magnetic moments, where the intra-layer interactions in the $ab$-plane are ferromagnetic and the inter-layer interaction along $c$-axis is antiferromagnetic.  Figure~\ref{Fig8}(b) shows CEF analysis of the isothermal magnetization at 2~K.  The small discrepancy between the calculated CEF curve and the experimental magnetization may be attributed to the fact that the CEF calculations do not take into account the Kondo effect.  However, the anisotropy in the magnetization plots is clearly explained semi-quantitatively by present set of crystal field parameters.

%*********************Table 2**************************
\begin{table*}[!]
\centering
\caption{CEF fit parameters, energy levels and wave functions}
\label{table2}
\begin{tabular}{ccccccc} 
\hline
\multicolumn{7}{l}{CEF Parameters}                                                                                                                                                                                                                                                                                                                                  \\ 
\hline
      & \begin{tabular}[c]{@{}c@{}}$B_2^0$ \\(K)\end{tabular} & \begin{tabular}[c]{@{}c@{}}$B_2^2$\\~(K)\end{tabular} & \begin{tabular}[c]{@{}c@{}}$B_4^0$ \\(K)\end{tabular} & \begin{tabular}[c]{@{}c@{}}$B_4^2$ \\(K)\end{tabular} & \begin{tabular}[c]{@{}c@{}}$B_4^4$ \\(K)\end{tabular} & \begin{tabular}[c]{@{}c@{}}$\lambda_{\rm i}$ \\(emu/mol)$^{-1}$\end{tabular}                         \\ 
\hline
      & $5.79$                                              & $-2.49$                                             & $-2.40$                                             & $-1.18$                                             & $1.74$                                              & \begin{tabular}[c]{@{}c@{}}$\lambda_{\rm x}$= 30\\ $\lambda_{\rm y}$ = 30\\$\lambda_{\rm z}$ = -25\end{tabular}  \\ 
\hline
\multicolumn{7}{c}{Energy levels and wave functions}                                                                                                                                                                                                                                                                                                                \\ 
\hline
\\
E (K) & $\ket{+ \frac{5}{2}}$                                       &  $\ket{+ \frac{3}{2}}$                                             &  $\ket{+ \frac{1}{2}}$                                                                                       &  $\ket{- \frac{1}{2}}$                                                                                     &  $\ket{- \frac{3}{2}}$                                             &  $\ket{- \frac{5}{2}}$                                                
 \\ \\
 \hline
 \\
767   & 0                                                 & 0.996                                             & 0                                                 & 0.014                                             & 0                                                 & 0.0901                                                                                  \\
767   & 0.0901                                            & 0                                                 & 0.014                                             & 0                                                 & 0.996                                             & 0                                                                                       \\
259   & 0                                                 & -0.0865                                           & 0                                                 & -0.167                                            & 0                                                 & 0.982                                                                                   \\
259   & 0.982                                             & 0                                                 & 0.167                                             & 0                                                 & 0.0865                                            & 0                                                                                       \\
0     & 0                                                 & -0.0288                                           & 0                                                 & 0.986                                             & 0                                                 & 0.165                                                                                   \\
0     & 0.165                                             & 0                                                 & 0.986                                             & 0                                                 & -0.0288                                           & 0                                                                                       \\
\hline
\end{tabular}
\end{table*}

%*********************************************************************

The value of crystal field split energy levels of the $2J+1$ ground state of the Ce-atom have been used to analyse the heat capacity data.  The $4f$ contribution of the heat capacity, in the paramagnetic state can be expressed as the contribution of the electronic, Kondo and the Schottky terms as given below:

\begin{equation}
\label{eqn9}
C_{\rm 4f} = C_{\rm el} + C_{\rm K} + C_{\rm Sch}.
\end{equation}

%*********************FIGURE 8********************************
\begin{figure}[!]
\includegraphics[width=0.5\textwidth]{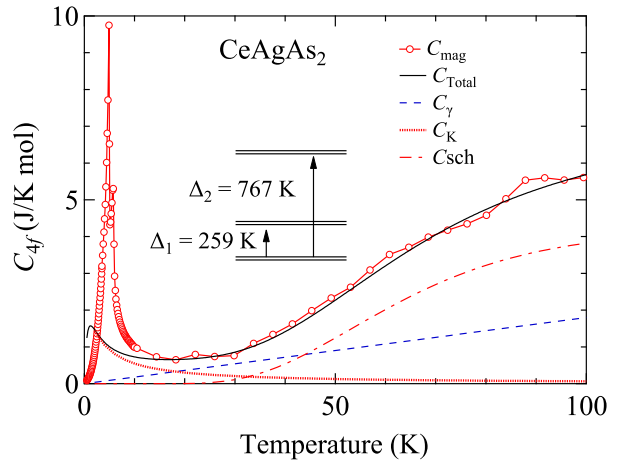}
\caption{\label{Fig9}(Color online) $4f$ contribution the heat capacity of CeAgAs$_2$.  The thick solid line is the calculated $C_{4f}$ based on Eqn.~\ref{eqn9}.  }
\end{figure}
%**************************************************************** 

The $C_{\rm el}$ is the electronic heat capacity defined by $\gamma T$, while the expressions for $C_{\rm K}$ and $C_{\rm Sch}$ are given in Ref~[\onlinecite{Arvind}]. The black solid line in Fig.~\ref{Fig8}, is the calculated curve based on Eqn.~\ref{eqn9}. The energy levels obtained from the CEF analysis of the magnetic susceptibility data, reproduce well the Schottky heat capacity. We obtain a Kondo temperature of 2.5~K. An estimation of the Kondo temperature in the mean-field model can also be obtained from the jump in the $C_{\rm 4f}$ of a Kondo system using the expression given by Blanco \textit{et al}~\cite{Blanco}.  The jump in the magnetic part of the heat capacity at $T_{\rm N2}$ amounts to 9.02~J/K$\cdot$mol, which results in a $T_{\rm K}/T_{\rm N}$ value of 0.35 and hence a  Kondo temperature of 1.72~K, which is close to the value obtained above using Eqn.~\ref{eqn9}.   Although, our estimation of the crystal field levels from the magnetic susceptibility data explains the Schottky heat capacity well which may be taken as a consistency check, for a precise determination of these levels, inelastic neutron diffraction has to be performed, which is planned as a future work.

\section{Summary}

We were successful in growing a single crystal of CeAgAs$_2$ and its non-magnetic analog LaAgAs$_2$ by flux method using Ag:As eutectic composition as flux.   From the x-ray diffraction and the TEM analysis we confirmed the superstructure in CeAgAs$_2$ single crystals which results in an orthorhombic variant of the HfCuSi$_2$ type tetragonal crystal structure, with two distinct crystallographic sites for the Ce-atom.  The RKKY-type interaction results in two magnetic transitions at $T_{\rm N1} = 6$~K and $T_{\rm N2} = 4.9$~K.  The magnetic susceptibility is highly anisotropic and it exhibits a broad hump in the temperature range 50 to 300~K for field parallel to [001] direction. From the analysis of magnetization based on point charge model of crystal electric field combined with the heat capacity data, it is confirmed that this broad hump is due to the crystal electric field and not due to the intermediate valence or mixed valence of Ce-atoms.     The anisotropy in the magnetic susceptibility can also be clearly explained by the crystal field analysis, with the two excited states at 259~K and 767~K, respectively. The magnetic easy axis is found to be along [100] or [010] direction, which is in conformity with the previous neutron diffraction studies. The negative logarithmic increase in the electrical resistivity at low temperature in the paramagnetic region together with the reduced magnetic moment and reduced heat capacity jump at the magnetic transition confirms the presence of Kondo interaction in this compound. 

\section{Acknowledgement}

We thank Mr. Jayesh Parmar and Ms. Ruta Kulkarni for their help in performing TEM and other measurements.  The discussion with Christoph Geibel, MPI-CPS, Dresden is gratefully acknowledged.

\end{document}